\documentclass[10pt]{article}

\usepackage{amsmath}
\usepackage{array}
\usepackage{appendix}
\usepackage{graphicx}
\usepackage{amsfonts}
\usepackage{amssymb}
\usepackage{mathrsfs}
\usepackage{yfonts}
\usepackage{euscript}
\usepackage{upgreek}
\usepackage{slantsc}
\usepackage{calligra}
\usepackage[T1]{fontenc}
\usepackage{epsf}
\usepackage{latexsym}

\usepackage{tipa}

\def\be{\begin{equation}}
\def\ee{\end{equation}}
\def\beq{\begin{equation}}
\def\eeq{\end{equation}}
\def\bea{\begin{eqnarray}}
\def\eea{\end{eqnarray}}

\def\ni{\noindent}

\def\!{\hspace{-1.6667em}}

\def\mF{\mbox{F}}

\def\mV{\mbox{V}}

\def\ttQ{\mbox{\tt Q}}

\def\brho{\mbox{\boldmath$\rho$}}          
\def\bomega{\mbox{\boldmath$\omega$}}      

\def\fA{\mbox{\sffamily A}}

\def\fF{\mbox{\sffamily F}}
\def\fG{\mbox{\sffamily G}}

\def\fI{\mbox{\sffamily I}}

\def\fP{\mbox{\sffamily P}}
\def\fQ{\mbox{\sffamily Q}}

\def\fS{\mbox{\sffamily S}}

\def\fX{\mbox{\sffamily X}}

\def\fZ{\mbox{\sffamily Z}}

\def\P{\underline{P}}

\def\bM{\mbox{\bf M}}

\def\bM{\mbox{{\bf M}}}
\def\bM{\mbox{{\bf M}}}
\def\bN{\mbox{{\bf N}}}

\def\scC{\mbox{\scriptsize ${\cal C}$}}          
\def\scE{\mbox{\scriptsize ${\cal E}$}}          
\def\scF{\mbox{\scriptsize ${\cal F}$}}
\def\scH{\mbox{\scriptsize ${\cal H}$}}          
\def\scI{\mbox{\scriptsize ${\cal I}$}}

\def\scL{\mbox{\scriptsize ${\cal L}$}}          
\def\scN{\mbox{\scriptsize ${\cal N}$}}
\def\scO{\mbox{\scriptsize ${\cal O}$}}
\def\scP{\mbox{\scriptsize ${\cal P}$}}
\def\scR{\mbox{\scriptsize ${\cal R}$}}          
\def\scS{\mbox{\scriptsize ${\cal S}$}}

\def\scU{\mbox{\scriptsize ${\cal U}$}}          

\def\iB{\mbox{\scriptsize$B$}}   

\def\FrQ{\mbox{\Large $\mathfrak{q}$}}

\def\FrC{\mbox{\Large $\mathfrak{c}$}}

\def\FrM{\mbox{\Large $\mathfrak{m}$}}                         

\def\sFG{\mbox{$\mathfrak{g}$}}

\def\FrG{\mbox{\Large $\mathfrak{g}$}}  
 
\def\sa{\mbox{\scriptsize a}}

\def\scc{\mbox{\scriptsize c}}
\def\sd{\mbox{\scriptsize d}}
\def\se{\mbox{\scriptsize e}}
\def\sf{\mbox{\scriptsize f}}
 
\def\sh{\mbox{\scriptsize h}} 
\def\si{\mbox{\scriptsize i}}

\def\sll{\mbox{\scriptsize l}}  
\def\sm{\mbox{\scriptsize m}}
\def\sn{\mbox{\scriptsize n}} 
\def\so{\mbox{\scriptsize o}}

\def\sr{\mbox{\scriptsize r}}
\def\st{\mbox{\scriptsize t}}

\def\sx{\mbox{\scriptsize x}}

\def\sB{\mbox{\scriptsize B}}

\def\sE{\mbox{\scriptsize E}}

\def\sJ{\mbox{\scriptsize J}}

\def\sM{\mbox{\scriptsize M}}

\def\sS{\mbox{\scriptsize S}}
\def\sT{\mbox{\scriptsize T}}

\def\sfz{\mbox{\sffamily{\scriptsize z}}}

\def\sfA{\mbox{\sffamily{\scriptsize A}}}      
\def\sfB{\mbox{\sffamily{\scriptsize B}}}      
\def\sfC{\mbox{\sffamily{\scriptsize C}}}      
\def\sfF{\mbox{\sffamily{\scriptsize F}}}      
\def\sfG{\mbox{\sffamily{\scriptsize G}}}      
\def\sfI{\mbox{\sffamily{\scriptsize I}}}      
\def\sfM{\mbox{\sffamily{\scriptsize M}}}      
\def\sfO{\mbox{\sffamily{\scriptsize O}}}      
\def\sfP{\mbox{\sffamily{\scriptsize P}}}      
\def\sfQ{\mbox{\sffamily{\scriptsize Q}}}      
\def\sfR{\mbox{\sffamily{\scriptsize R}}}      
\def\sfS{\mbox{\sffamily{\scriptsize S}}}      
\def\sfV{\mbox{\sffamily{\scriptsize V}}}      
\def\sfW{\mbox{\sffamily{\scriptsize W}}}      
\def\sfX{\mbox{\sffamily{\scriptsize X}}}      
\def\sfY{\mbox{\sffamily{\scriptsize Y}}}      
\def\sfZ{\mbox{\sffamily{\scriptsize Z}}}      

\def\te{\mbox{\tiny e}}

\def\tm{\mbox{\tiny m}}

\def\tJ{\mbox{\tiny J}}

\def\tfA{\mbox{\sffamily{\tiny A}}}

\def\Q{\overline{\ttQ}}

\def\K{Kucha\v{r} }

\def\pa{\partial}
\def\d{\textrm{d}}

\def\Last{\mbox{\Large$\ast$}}                
\def\5Star{\mbox{\Large$\star$}}              
\def\Ast{\mbox{\Large$\ast$}}                 %

\def\cr{\mbox{\scriptsize{\bf $\mbox{ } \times \mbox{ }$}}}

\def\sumi2{\sum\mbox{}_{\mbox{}_{\mbox{\scriptsize $i$=1}}}^2}
\def\sumi3{\sum\mbox{}_{\mbox{}_{\mbox{\scriptsize $i$=1}}}^3}

\def\sumj3{\sum\mbox{}_{\mbox{}_{\mbox{\scriptsize $j$=1}}}^3}
\def\sumk3{\sum\mbox{}_{\mbox{}_{\mbox{\scriptsize $k$=1}}}^3}

\begin{document}

\begin{titlepage}

\begin{center}

\Huge{\bf TRiPoD}\normalsize

\vspace{0.08in}

{\bf \Large(Temporal Relationalism incorporating Principles of Dynamics)}

\vspace{.15in}

{\large \bf Edward Anderson} 

\vspace{.15in}

\large {\em DAMTP, Centre for Mathematical Sciences, Wilberforce Road, Cambridge CB3 OWA.  } \normalsize

\end{center}

\begin{abstract}
 
Temporal Relationalism is that there is no time for the universe as a whole at the primary level. 
Time emerges rather at a secondary level; one compelling idea for this is Mach's: that time is to be abstracted from change.
Temporal Relationalism leads to, and better explains, the well-known Frozen Formalism Problem encountered in GR and other background-independent theories at the quantum level.
Abstraction from change is then a type of emergent time resolution of this. 
Moreover, the Frozen Formalism Problem is but one of the many Problem of Time facets, which are notoriously interconnected.
The current article concerns modifications of physical formalism which ensure that once Temporal Relationalism is resolved, it stays incorporated.
At the classical level, this involves modifying much of the Principles of Dynamics.

I first introduce the anti-Routhian to complete the Legendre square of Lagrangian, Hamiltonian and Routhian.   
I next pass from velocities $\dot{Q}^{\sfA}$ to changes $\d Q^{\sfA}$.  
Then Lagrangians are supplanted by Jacobi arc elements, Euler--Lagrange equations by Jacobi--Mach ones, and momentum requires redefining but actions remain unchanged. 
A differential (d) version of the Hamiltonian is required, giving rise to a variant of the Dirac approach based on a d-almost Hamiltonian subcase of the d-anti Routhian.
On the other hand, the forms of the constraints themselves, and of Hamilton--Jacobi theory, remain unaltered.

\end{abstract}

\section{Introduction}\label{Introduction}

\ni {\it Temporal Relationalism} is that there is no time for the universe as a whole at the primary level \cite{L}.
This can be implemented by a formulation of the following kind.  

\mbox{ } 

\ni i)  It is not to include any appended times -- such as Newtonian time -- or appended time-like variables. 

\ni ii) Time is not to be smuggled into the formulation in the guise of a label either.

\mbox{ }

\ni To formulate this, we need to know what {\sl does} remain available, which starts with the configurations $Q^{\sfA}$.
The collection of all possible values of the configuration of a physical system form the configuration space $\FrQ$; $k := \mbox{dim}(\FrQ)$.

\mbox{ } 

\ni Then one implementation of ii) is for a label to be present but physically meaningless because it can be changed for any other (monotonically related) 
label without changing the physical content of the theory.   
E.g. at the level of the action, this is to be {\it Manifestly Reparametrization Invariant}:   

\ni\beq
S = 2 \int \d\lambda \sqrt{T W} \mbox{ } , \mbox{ } \mbox{ for } \mbox{ } 
T := ||\dot{\mbox{\boldmath{$Q$}}}||_{\mbox{\scriptsize\boldmath{$M$}}}\mbox{}^2/2 := M_{\sfA\sfA^{\prime}}\dot{Q}^{\sfA}\dot{Q}^{\sfA^{\prime}}/2 \mbox{ } .  
\label{S-Rel}
\eeq
Therein, $T$ is the the {\it kinetic energy}, $\mbox{\boldmath{$M$}}$ the {\it configuration space metric} and $W = W(\mbox{\boldmath{$Q$}})$ is the {\it potential factor}.\footnote{In 
this Article, I often use particular sans serif indices for particular Principles of Dynamics (PoD) objects, then using primed versions when more than one index of a given kind is required.
I use bold font for coordinate-free version in configuration space and other spaces familiar from the PoD, and underlines for spatial vectors.
Particle number indices $I, J$ run from $1$ to $N$, relative inter-particle cluster indices $A = 1$ to $N - 1$, spatial dimension indices $i$ run from 1 to $d$. 
$m_I$ are the particles' masses.}

\mbox{ } 

\ni Arena 1) Jacobi's action principle \cite{Lanczos, Arnol'd}, which can be considered to be for temporally-relational spatially-absolute Mechanics.  
Here $T = ||\dot{\mbox{\boldmath$q$}}||_{\mbox{\scriptsize\boldmath $m$}}\mbox{}^2/2 = m_I\dot{q}^{iI}\dot{q}^{iI}/2$ so the configuration space metric $\mbox{\boldmath $m$}$ is just the 
`mass matrix' with components $m_{I}\delta_{IJ}\delta_{ij}$, and $W = E - V(\mbox{\boldmath $q$})$ for $V(\mbox{\boldmath{$q$}})$ the potential energy and $E$ is the total energy of the 
model universe.  

\mbox{ }

\ni Arena 2) Scaled 1-$d$ RPM with translations trivially removed \cite{FileR} is another case of Jacobi's action principle.
Here  $T = ||\dot{\brho}||^2/2$ and $W = E - V(\brho)$ for $\brho$ the mass-weighted relative Jacobi coordinates with components $\rho^{iA}$ 
\cite{Marchal, LR97, FileR}.
This has the advantage over 1) of being a relational whole-universe model. 

\mbox{ }

\ni Arena 3) Full GR can indeed also be formulated in this manner.  
This involves an increasing number of departures of formalism from the familiar Arnowitt--Deser--Misner (ADM) action \cite{ADM}, though all of these formalisms remain equivalent.
For instance, the ADM lapse of GR is an appended time-like variable.
It can however be removed in the Baierlein--Sharp--Wheeler (BSW) action \cite{BSW, B94I}, 
and then the ADM shift can be replaced with various distinct but equivalent auxiliaries to match the rest of each relational formalism \cite{RWR, FEPI, ARel, AM13}.
Because this example strays outside of the current Article's account of finite models, we do not elaborate on its details, 
though here the configurations are spatial 3-metrics and the configuration space metric is the inverse of the DeWitt supermetric \cite{DeWitt67}, which is indefinite.

\mbox{ }

\end{titlepage}

\ni Arena 4) Misner's action principle \cite{Magic} is for the finite minisuperspace subcase of the previous example. 
This already exhibits the indefinite configuration space metric feature. 
The full GR potential factor, $R - 2\Lambda$ for $R$ the spatial Ricci scalar and $\Lambda$ the cosmological constant, carries over straightforwardly to the minisuperspace case. 

\mbox{ } 

\ni The implementation of Temporal Relationalism can be further upgraded as follows.
It is a further conceptual advance to formulate one's action and subsequent equations without use of any meaningless label at all.  
I.e. a {\it Manifestly Parametrization   Irrelevant} formulation in terms of {\it change}           $\d Q^{\sfA}$             rather than 
     a      Manifestly Reparametrization Invariant           one in terms of a label-time velocity  $\d Q^{\sfA}/\d \lambda$.  

\mbox{ } 
	 
\ni However, it is better still to formulate this directly: without even mentioning any meaningless label or parameter, 
by use of how the preceding implementation is dual to a Configuration Space Geometry formulation. 
Indeed, Jacobi's action principle is often conceived of geometrically \cite{Lanczos, Arnol'd} terms rather than in its dual aspect as a timeless formulation.
This formulation's action now involves not kinetic energy $T$ but a kinetic arc element $\d s$: 

\ni\beq
S = \sqrt{2} \int \d s \sqrt{W} \mbox{ } , \mbox{ } \mbox{ } 
\d s := ||\d \mbox{\boldmath$Q$}||_{\mbox{\scriptsize\boldmath$M$}} = \sqrt{M_{\sfA\sfB}(\mbox{\boldmath$Q$})\d Q^{\sfA}\d Q^{\sfB}} \mbox{ } . 
\label{S-Rel-2}
\eeq
The above two forms of Temporal Relationalism implementing actions are clearly equivalent to each other.
See e.g. \cite{Lanczos, Arnol'd, BSW, RWR, FEPI, FileR, AM13} for how the product-type actions (\ref{S-Rel}, \ref{S-Rel-2}) 
are furthermore equivalent to the more familiar difference-type actions of Euler--Lagrange for Mechanics and of ADM for GR.  

\mbox{ }  

\ni The main way in which actions implementing ii) work is that they necessarily imply {\it primary constraints}.  
I.e. relations between the momenta that are obtained without use of the equations of motion; see Sec \ref{Constraints} for more context.
For (\ref{S-Rel}), this implication is via the following well-known argument of Dirac \cite{Dirac}. 
An action that is reparametrization-invariant is homogeneous of degree 1 in the velocities.  
Thus the $k$ conjugate momenta are (by the above definition) homogeneous of degree 0 in the velocities. 
Therefore they are functions of at most $k - 1$ ratios of the velocities. 
So there must be at least one relation between the momenta themselves (i.e. without any use made of the equations of motion).  
But this is the definition of a primary constraint.   
In this manner, Temporal Relationalism acts as a constraint provider.

The constraint it provides has a purely quadratic form induced from \cite{B94I} that of the action (\ref{S-Rel-2})\footnote{`Purely' here  
means in particular that there is no accompanying linear dependence on the momenta.  
At the level of conic sections and higher-dimensional quadratic forms, this means a `centred' choice of coordinates, meaning that the conic section is centred about the origin.} 
\beq
\scC\scH\scR\scO\scN\scO\scS :=  N^{\sfA\sfB} P_{\sfA} P_{\sfB}/2 - W( \mbox{\boldmath$Q$}) = 0   \mbox{ }  
\label{CHRONOS}
\eeq
for $\bN$ the inverse of the configuration space metric $\bM$. 
For instance, for Mechanics it is of the form $\scE := ||\mbox{\boldmath{$p$}}||_{\mbox{\scriptsize\boldmath{$n$}}} + V(\mbox{\boldmath{$q$}}) = E$, for 
$\mbox{\boldmath{$n$}} = \mbox{\boldmath{$m$}}^{-1}$, with components $\delta_{IJ}\delta_{ij}/m_I$.  
This usually occurs in Physics under the name and guise of an {\it energy constraint}, though this is not the interpretation it is afforded in the relational approach.
On the other hand, for GR, it is the well-known and similarly quadratic Hamiltonian constraint $\scH$ -- now containing the DeWitt supermetric itself -- 
that arises at this stage as a primary constraint.  

\mbox{ } 

\ni Reconciling timelessness for the universe as a whole at the primary level with time being apparent none the less in the parts of the universe that we observe,
proceeds via Mach's Time Principle \cite{M}: that `time is to be abstracted from change'. 
I.e. a particular type of emergent time at the secondary level.
This Machian position is particularly aligned with the second and third formulations of Temporal Relationalism ii), which are indeed in terms of change rather than velocity.
Machian times are of the general form
\beq  
t^{\se\sm(\sM\sa\scc\sh)} = F[  Q^{\sfA}, \d  Q^{\sfA}] \mbox{ } . 
\label{t-Mach}
\eeq
\mbox{ }\mbox{ } More specifically, \cite{B94I, FileR, ARel2, ABook} one is best served by adopting a conception of time along the lines of the astronomers' ephemeris time \cite{Clemence}.
This is from including a sufficient totality of locally relevant change. 

\mbox{ } 

\ni A specific implementation of this comes from rearranging (\ref{CHRONOS}).
This amounts to interpreting this not as an energy-type constraint but as an {\it equation of time} thus called $\scC\scH\scR\scO\scN\scO\scS$ (after the primordial Greek God of Time). 
We shall see that this rearrangement is aligned with 

\ni\beq 
\Ast := \frac{\pa}{\pa t^{\te\tm(\tJ)}} = \frac{\sqrt{2W}}{\d s}    \frac{\pa}{\pa\lambda}
\eeq 
simplifying\footnote{Choosing 
time so that the equations of motion are simple was e.g. already argued for in \cite{MTW}; see also Sec 3.4.}
the system's momenta and equations of motion. 
Integrating up, 
\beq
t^{\se\sm(\sJ)} - t^{\se\sm(\sJ)}(0)  = \int \d s/\sqrt{2W}  \mbox{ } \mbox{ {\it (emergent Jacobi time) }} . 
\label{t-em-J}
\eeq

\mbox{ }  

\ni At the quantum level, (\ref{CHRONOS}) gives rise to   
\beq
\widehat{\scC\scH\scR\scO\scN\scO\scS}\Psi = 0 \mbox{ } ,
\label{Q-CHRONOS}
\eeq
for $\Psi$ the wavefunction of the universe.
This includes the time independent Schr\"{o}dinger equation $\widehat{\scE}\Psi = E\Psi$ for Mechanics 
and the well-known Wheeler--DeWitt equation  \cite{DeWitt67, Battelle} $\widehat{\scH}\Psi = 0$. 
Moreover, these equations occur in a situation in which one might expect time-dependent Schr\"{o}dinger equations $\widehat{H}\Psi = i\hbar\pa\Psi/\pa t$ for some notion of time $t$.
Thus (\ref{Q-CHRONOS}) exhibits a {\it Frozen Formalism Problem} \cite{DeWitt67, Kuchar92I93}.         
This is a well-known facet of the Problem of Time (PoT); the preceding steps which trace this back to the classical level and point to Machian resolutions are for now less well-known.

\mbox{ } 

\ni Moreover, the above suite of implementations of Temporal Relationalism can be extended to cover {\sl every} physical formalism, 
rather than just at the level of the Lagrangian (or Jacobi) formulations.
This is necessary in order for Temporal Relationalism to be tractable in tandem with the many other facets of the PoT \cite{Kuchar92I93, APoTAPoT2, FileR, APoT3, ABook}.
Relevant parts of the standard PoD are recollected in Sec 2 and extended by completion of the square of Legendre transformations to include `anti-Routhians'. 
This Article concentrates on finite models, though its results readily carry over to field theory also (including in particular full GR) \cite{ARel, FileR, ABook}.

The idea now is to compose Temporal Relationalism with the other PoT facets (see \cite{APoT3} for a recent account with facets specifically laid out piecemeal out prior to composition).
{\sl Reformulating the PoD to be Temporal Relationalism incorporating -- TRiPoD -- ensures that all subsequent considerations of other Problem of Time facets 
within this new paradigm do not violate the Temporal Relationalism initially imposed} (Sec 3).\footnote{I subsequently use the acronym TRi more widely in this Article.}
This is a big issue because the PoT facets have been likened to the gates of an enchanted castle \cite{Kuchar93}: if one goes through one of them and then through another, 
one often finds oneself once again outside of the first one.
In the current Article I reformulate the entirety of the PoD to ensure that one stays within Temporal Relationalism in one's subsequent efforts to pass through further PoT facets' gates.
This is indicatory that a lot of work is required in order to not lose one's program's previous successes 
with a subset of the gates when one tries to extend that program to pass through more of the gates. 
On the other hand, that this can be done at all is encouraging as a sign of the non-impossibility of the whole (or at least local) PoT.

Thus I next pass from velocities $\dot{Q}^{\sfA}$ to changes $\d Q^{\sfA}$ to align with equations (2), (4) and (5).   
Then Lagrangians are supplanted by Jacobi arc elements, Euler--Lagrange equations by Jacobi--Mach ones, and momentum requires redefining but actions remain unchanged. 
The TRiPoD version of the PoD is similar in spirit to Dirac's introduction of multiple further notions of Hamiltonian to start to deal systematically with constrained systems, 
by appending diverse kinds of constraints with diverse kinds of Lagrange multipliers.
A differential (d) version of the Hamiltonian is required to be TRi.  
This affects such as the total Hamiltonian, which is now approached instead by appending cyclic differentials\footnote{To avoid confusion, 
note that `cyclic' in `cyclic differential' just means the same as `cyclic' in cyclic velocity, rather than implying some particular kind of differential itself. 
Thus nothing like `exact differential' or `cycle' in algebraic topology -- which in de Rham's case is tied to differentials -- is implied.} 
rather than Lagrange multipliers.
This gives rise to a variant of the Dirac approach based on a d-almost Hamiltonian subcase of d-anti Routhian.
On the other hand, the forms of the constraints themselves are unaltered, 
as are the constraint algebraic structure and expression in terms of beables are also unaltered in the finite-model case.  
Finally, Hamilton--Jacobi theory remains unaltered.

This Article concludes in Sec 4 by briefly pointing out the TRi-Foliations \cite{TRiFol} and TRi-canonical quantum mechanics extensions that complete the TRi program  
within which to tackle the rest of the PoT.

\section{The Standard Principles of Dynamics}\label{PoD}

\subsection{Lagrangians and Euler--Lagrange Equations}\label{Lagrangian}

Begin by considering a for now finite second-order classical physical system \cite{Lanczos, Goldstein} expressed in terms of Lagrangian variables $Q^{\sfA}, \dot{Q}^{\sfA}$.  
All dynamical information is contained within the {\it Lagrangian} function $L(Q^{\sfA}, \dot{Q}^{\sfA}, t)$.  
The commonest form for this is $L = T - V(\mbox{\boldmath $Q$}, t)$ for $T$ as in (\ref{S-Rel}) but with the dot signifying $\pa/\pa t$ rather than 
$\pa/\pa\lambda$.\footnote{Note that this Article 
concentrate on systems with no extra explicit time dependence (currently in the Lagrangian).}
%
Then apply the standard prescription of the Calculus of Variations to obtain the equations of motion such that the action $S = \int \d t \, L$ is stationary with respect to $Q^{\sfA}$.
This approach considers the true motion between two particular fixed endpoints $e_1$ and $e_2$ along with the set of varied paths (subject to the same fixed endpoints) about this motion.  
This gives the {\it Euler--Lagrange equations} 
\be
\frac{\d }{\d t} \left\{ \frac{\pa L}{\pa \dot{Q}^{\sfA}} \right\} - \frac{\pa L}{\pa Q^{\sfA}} = 0   \mbox{ }. 
\label{ELE-2}
\ee
\mbox{ } \mbox{ } These equations simplify in the below three special cases, two of which involve particular types of coordinates.
Indeed, one major theme in the principles of dynamics is judiciously chosen a coordinate system with as many simplifying coordinates as possible.  

\mbox{ }

\ni 1) {\it Lagrange multiplier coordinates} $m^{\sfM} \, \underline{\subset} \, Q^{\sfA}$ are such that $L$ is independent of $\dot{m}^{\sfM}$: 
\beq
\pa L/\pa\dot{m}^{\sfM} = 0 \mbox{ } .
\label{prelim-m}
\eeq
Then the corresponding Euler--Lagrange equation simplifies to  
\be
\pa L/\pa m^{\sfM} = 0 \mbox{ } .
\label{lmel}
\ee
2){\it Cyclic coordinates} $c^{\sfY} \, \underline{\subset} \, Q^{\sfA}$  are such that $L$ is independent of $c^{\sfY}$:
\beq
\pa L/\pa c^{\sfY} = 0 \mbox{ } ,
\eeq 
but features $\dot{c}^{\sfY}$: the corresponding {\it cyclic velocities}.
Then the corresponding Euler--Lagrange equation simplifies to 
\be
\pa L/\pa \dot{c}^{\sfY} = \mbox{ const}^{\sfY} \mbox{ } .
\label{cyclic-vel}
\ee
3){\it The energy integral type simplification}.  
If $L$ is independent of the independent variable itself (usually $t$): $\pa L/\pa t = 0$, then one Euler--Lagrange equation may be supplanted by the first integral
\be
L - \dot{Q}^{\sfA} \frac{\pa L}{\pa \dot{Q}^{\sfA}} = \mbox{ constant } \mbox{ } .
\label{en-int}
\ee
Further suppose that 1)'s equations 
\be
0 = \frac{\pa L}{\pa m^{\sfM}}(Q^{\sfO}, \dot{Q}^{\sfO}, m^{\sfM}) \mbox{ happen to be solvable for } m^{\sfM} \mbox{ } .
\eeq
(here $Q^{\sfO}$ denotes `other than multiplier coordinates').

Then one can pass from $L(Q^{\sfO}, \dot{Q}^{\sfO}, m^{\sfM})$ to a reduced $L_{\sr\se\sd}(Q^{\sfO}, \dot{Q}^{\sfO})$; this is known as {\it multiplier elimination}.

\subsection{Conjugate momenta}

These are defined by
\be 
P_{\sfA} := \pa L/\pa\dot{Q}^{\sfA} \mbox{ } . 
\label{mom-vel}
\ee
Explicit computation of this then gives the {\it momentum--velocity relation}
\be
P_{\sfA} = M_{\sfA\sfB}\dot{Q}^{\sfB} \mbox{ } .
\ee
\mbox{ } \mbox{ } As a first application of this, it permits rewriting some of the preceding simplifications. 
The preliminary condition (\ref{prelim-m}) in deducing the multiplier condition is now $\dot{P}^{\sfY} = 0$.
The cyclic coordinate condition (\ref{lmel}) is now
\be
P^{\sfY} = \mbox{ const}^{\sfY} \mbox{ } ,
\label{cyclic-vel-P}
\ee
whereas the energy integral (\ref{en-int}) is
\beq
L - \dot{Q}^{\sfA}P_{\sfA} = \mbox{ constant } \mbox{ } .
\eeq

\subsection{Legendre transformations}\label{Legendre}

Suppose one has a function $F(x_{\sfW}, v_{\sfV})$ and one wishes to use $z_{\sfW} = \pa F/\pa x_{\sfW}$ as variables in place of the $x_{\sfW}$.  
To avoid losing information in the process, a {\it Legendre transformation} is required, by which one passes to a function  
\beq
\mbox{$G(z_{\sfW}, v_{\sfV}) = x_{\sfW}z_{\sfW} - F(x_{\sfW}, v_{\sfV})$ } .
\eeq  
Legendre transformations are symmetric between $x_{\sfW}$ and $z_{\sfW}$: if one defines $x_{\sfW} := \pa G/\pa z_{\sfW}$, 
the reverse passage is now to $F(x_{\sfW}, v_{\sfV}) = x_{\sfW}z_{\sfV} - G(z_{\sfW}, v_{\sfV})$.  

\mbox{ }

\ni In particular, in Mechanics if one's function is a Lagrangian $L(Q_{\sfA}, \dot{Q}^{\sfA})$, one may wish to use some of the conjugate momenta $P_{\sfA}$
as variables in place of the corresponding velocities $\dot{Q}^{\sfA}$.  

\mbox{ } 

%
\ni Example 1) {\it Passage to the Routhian}.
Given a Lagrangian with cyclic coordinates $c^{\sfY}, \mbox{ } L(Q^{\sfX}; \dot{Q}^{\sfX}, \dot{c}^{\sfY})$, then 

\ni $\pa L / \pa\dot{c}^{\sfY} := P_{\sfY} = const^{\sfY}$. 
Thus one may pass from $L$ to the {\it Routhian} 
\beq
R(Q^{\sfX}, \dot{Q}^{\sfX}, p^c_{\sfY}, t) := L(Q^{\sfX}, \dot{Q}^{\sfX}, \dot{c}^{\sfY}, t) - P^c_{\sfY}\dot{c}^{\sfY} \mbox{ } .
\label{Routhian}
\eeq  
This amounts to treating the cyclic coordinates, as a package, differently from the non-cyclic ones.

\mbox{ } 

\ni Much of the motivation for the Routhian is that it is in some ways a useful trick to treat cyclic coordinates differently from the others. 
The usual setting for this is simplifying the Euler--Lagrange equations. 
This requires the cyclic velocities analogue of multiplier elimination, termed {\it Routhian reduction}, which requires being able to solve
\beq
{\it const}_{\sfY} = \frac{\pa L}{\pa \dot{c}^{\sfY}}(Q^{\sfX}, \dot{Q}^{\sfX}, \dot{c}^{\sfY}) \mbox{ as equations for the } \dot{c}^{\sfY} \mbox{ } . 
\eeq
Moreover, unlike in the corresponding multiplier elimination, one does not just substitute these back into the Lagrangian.
Rather, one aditionally needs to apply the Legendre transformation in (\ref{Routhian}), thus indeed passing to a Routhian rather than just a reduced Lagrangian.  
If the above reduction can be performed, then one can furthermore use the status of the cyclic momenta as constants (\ref{cyclic-vel-P}) 
to end any further involvement of the cyclic variables in the dynamical problem at hand.
I.e. the corresponding part of the integration of equations of motion has thereby been completed.
Note finally that the manoeuvre \cite{Lanczos} from Euler--Lagrange's action principle with no explicit $t$ dependence to Jacobi's action principle is a subcase of Routhian reduction.   

\mbox{ }

\ni Example 2) {\it Passage to the Hamiltonian}. In general,
\beq
H(Q^{\sfA}, P_{\sfA}, t) := P_{\sfA}\dot{Q}^{\sfA} - L(Q^{\sfA}, \dot{Q}^{\sfA}, t) \mbox{ } ,
\eeq
which makes use of {\sl all} the conjugate momenta.    
This Article then concentrates on the $t$-independent version.
For instance, for the most common form of Lagrangian, 
\beq
H = ||\mbox{\boldmath{$P$}}||_{\mbox{\scriptsize\boldmath $N$}}\mbox{}^2/2 + \mV(\mbox{\boldmath{$Q$}}, t) \mbox{ }  \mbox{ } .
\eeq
In the general case, the equations of motion are {\it Hamilton's equations}, 
\be
\dot{Q}^{\sfA} = \pa H/\pa P_{\sfA} \mbox{ } , \mbox{ } \mbox{ } \dot{P}_{\sfA} = - \pa H/\pa Q^{\sfA} \mbox{ } .
\ee
In the time-dependent case, these are supplemented by $- \pa L/\pa t = \pa H/\pa t$.

For the Lagrange multiplier coordinates, half of the corresponding Hamilton's equations collapse to just 
\beq
\pa H/\pa m^{M} = 0 \mbox{ }   
\eeq
in place of (\ref{lmel})
On the other hand (\ref{en-int}) becomes $H = const$ in the time-dependent case.

The Hamiltonian formulation is further motivated firstly by admitting a systematic treatment of constraints due to Dirac (\cite{Dirac, HTbook} and Sec \ref{Constraints}). 
Secondly, it offers a more direct link to quantum theory.
N.B. also that in this case (within for the range of theories considered in this Article) 
the Legendre transformation is from a geometrical perspective effectuating passage from the tangent bundle $T(\FrQ)$ to the cotangent bundle $T^*(\FrQ)$.  

\mbox{ } 

\ni Example 3) The next Sec motivates consideration also of {\it passage to the anti-Routhian}
\beq
A(Q^{\sfX}, P^{\sfX}, \dot{c}^{\sfY}, t) := L(Q^{\sfX}, \dot{Q}^{\sfX}, \dot{c}^{\sfY}, t) - P_{\sfX}\dot{Q}^{\sfX} \mbox{ } .
\eeq
\mbox{ } \mbox{ } Introducing $A$ is natural insofar as it completes the Legendre square whose other vertices are $L$, $H$ and $R$ (Fig \ref{PoD-Squares}). 
Returning to the motivation of the Routhian itself, passage to the anti-Routhian turns out to be a distinct useful trick involving treating cyclic coordinates 
differently from the others, but now under the diametrically opposite Legendre transformation.
The smaller part of the new trick is in Sec \ref{Constraints}, whilst the larger part has to await the further special case that Sec 3 centres upon.

Moreover, both of these tricks in general come at a price. 
The first part of this price is geometrical, namely that these now involve slightly more complicated {\it mixed cotangent--tangent bundles} over $\FrQ$.  
I.e. $T(\check{\FrQ}) \times T^*(\FrC)$ for the Routhian case and $T^*(\check{\FrQ}) \times T(\FrC)$ for the anti-Routhian case, 
where $\FrC$ is the subconfiguration space of cyclic coordinates and $\check{\FrQ}$ is the complementary subconfiguration space of the $Q^{\sfX}$.   

{            \begin{figure}[ht]
\centering
\includegraphics[width=1.0\textwidth]{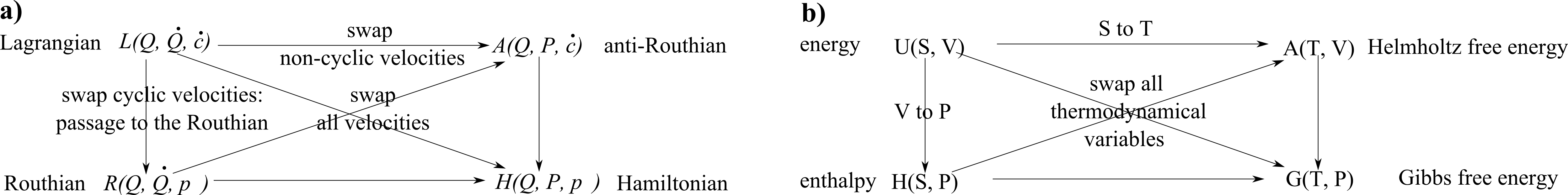}
\caption[Text der im Bilderverzeichnis auftaucht]{         \footnotesize{a) completes an elsewise well-known Legendre square by introducing an anti-Routhian 
that is the diametric opposite of the Routhian in terms of which variables it swaps by Legendre transformation.
I.e. it makes the same cyclic-other packaging, but converts {\sl the other set's} velocities to momenta.  
b) The analogous Legendre square from Thermodynamics, where all 4 corners are well-known: internal energy $U$, enthalpy $H$, Gibbs free energy $G$ and Helmholtz free energy $A$.} }
\label{PoD-Squares}\end{figure}            }

\subsection{Corresponding morphisms}\label{Mor}

The morphisms corresponding to $\FrQ$'s are the {\it point transformations} Point($\FrQ$): the coordinate transformations of $\FrQ$.  
These readily induce the transformation theory for Lagrangian variables.
More generally, but not in this Article, one has the time-dependent extension Point$_t$.  
The transformation theory for Hamiltonian variables is in general more subtle. 
This in part reflects that $P_{\sfA}\dot{Q}^{\sfA}$ becomes involved due to featuring in the conversion from $L$ to $H$. 
Here starting from Point, the momenta then follow suit by preserving $P_{\sfA}\dot{Q}^{\sfA}$ \cite{Lanczos}; these transformations still preserve $H$.
On the other hand, starting from Point$_t$ induces gyroscopic corrections to $H$ \cite{Lanczos}; this illustrates that $H$ itself can change form.
More general transformations which mix the $Q_{\sfA}$ and the $P^{\sfA}$ are additionally possible, 
though they are not as unrestrictedly general functions of their $2k$ arguments as Point's transformations are as functions of their $k$ arguments.
A first case are the transformations which preserve the {\it Liouville 1-form} 
\beq
P_{\sfA}\d Q^{\sfA}
\label{Liouville}
\eeq
that is clearly associated with $P_{\sfA}\dot{Q}^{\sfA}$.
These can again be time-independent (termed {\it scleronomous}) or time-dependent in the sense of parametrization adjunction of $t$ to the $Q^{\sfA}$ (termed {\it rheonomous}). 
Again, the former preserve $H$ whereas the latter induce correction terms \cite{Lanczos}.  
These are often known as {\it contact transformations}, so I denote them by Contact and Contact$_t$ respectively.  
Point, Point$_t$, Contact and Contact$_t$ form a diamond of subgroups.

More generally still, it turns out to be the integral of (\ref{Liouville}) that needs to be preserved.
Correspondingly, at the differential level, this means that (\ref{Liouville}) itself is preserved up to an additive complete differential $\d G$ for $G$ the {\it generating function}.
In this generality, one arrives at the {\it canonical transformations} alias {\it symplectomorphisms}, 
once again in the form of a rheonomous group Can$_t$ and a scleronomous subgroup Can.
Whilst arbitrary canonical transformations do not permit explicit representation, the infinitesimal ones do.  
Finally, applying Stokes' Theorem to the integral of (\ref{Liouville}) reveals a more basic invariant: the bilinear antisymmetric {\it symplectic 2-form} 
\beq
\d P_{\sfA} \wedge \d Q^{\sfA}
\label{Symplectic}
\eeq
(denoted by $\bomega$ with components $\omega_{\sfA\sfA^{\prime}}$ to cover a wider generality of cases \cite{Arnol'd}).  
Thus one arrives at brackets built from this, of which the Poisson bracket is the most important example. 

\mbox{ } 

%
\ni Concentrating on the $t$-independent case that is central to this Article, for the Routhian, the morphisms are 

\ni Point($\check{\FrQ}$) $\times$ Can($\FrC$), though the latter piece is usually ignored out of the $c^{\sfY}$ being absent and the $p_{\sfY}^c$ being constant.
For the $t$-independent anti-Routhian, the morphisms are Can($\check{\FrQ}$) $\times$ Point($\FrC$).
The above more complicated spaces are the second price to pay.

\subsection{Bracket structures}\label{Brackets}

The {\it Poisson bracket} $\mbox{\bf \{} \mbox{ } \mbox{\bf ,} \mbox{ }  \mbox{\bf \}}$ of quantities $F(Q^{\sfA}, P_{\sfA})$ and $G(Q^{\sfA}, P_{\sfA})$ is given by
\be
\mbox{\bf \{} F \mbox{\bf ,} \, G \mbox{\bf \}} := \frac{\pa F}{\pa Q^{\sfA}}\frac{\pa G}{\pa P_{\sfA}}     -   \frac{\pa G}{\pa Q^{\sfA}}\frac{\pa F}{\pa P_{\sfA}} \mbox{ } . 
\label{PB}
\ee
Then in terms of these, the equations of motion take the form 
\beq
\mbox{\bf \{} P_{\sfA} \mbox{\bf ,} \, H \mbox{\bf  \}} = \dot{P}_{\sfA} \mbox{ } , \mbox{ } \mbox{ } \mbox{\bf \{} Q^{\sfA} \mbox{\bf ,} \, H \mbox{\bf \}} = \dot{Q}^{\sfA} \mbox{ } . 
\eeq
Furthermore, for any $f(Q^{\sfA}, P_{\sfA},t)$, the total derivative $\d f/\d t = \mbox{\bf \{} f \mbox{\bf ,} \, H \mbox{\bf  \}} + \pa f/\pa t$, 
so if $f$ does not depend explicitly on $t$, the intuitive conserved quantity condition $0 = \d f/\d t$ becomes
\beq
\mbox{\bf \{} f \mbox{\bf ,} \, H \mbox{\bf  \}} = 0 \mbox{ } .  
\eeq

{\it Phase space} is the space $T^*(\FrQ)$ of both the $Q^{\sfA}$ and the $P_{\sfA}$ as equipped with $\mbox{\bf \{} \mbox{ } \mbox{\bf ,} \mbox{ }  \mbox{\bf \}}$.  

\mbox{ }  

\ni One can also at this stage go back and find an equivalent description to the Poisson bracket at the Lagrangian tangent bundle 
level rather than requiring the Hamiltonians cotangent bundle canonical variables $\mbox{\bf \{}q \mbox{\bf ,} \,  p \mbox{\bf \}}$ to be defined in advance.
The resulting notion is the {\it Peierls bracket} \cite{PeierlsEMS02DeWittBook}.  
A price to pay in using this rather than the Poisson bracket is that it involves rather more complicated mathematics in terms of Green's functions.
[Its explicit form is not required in this Article.]

\mbox{ } 

\ni Finally, the second part of Sec 2.3's price to pay if one uses a Routhian or anti-Routhian is that the mixed cotangent--tangent bundle nature of the variables further implies 
in general the involvement of {\it mixed Poisson--Peierls brackets}.

\subsection{Hamilton--Jacobi Theory}\label{HJT}

If one replaces $P_{\sfA}$ by ${\pa S}/{\pa Q^{\sfA}}$ in $H$, one obtains the {\it Hamilton--Jacobi equation}, whose most general form is  
\be
\pa S/\pa t + H(Q^{\sfA}, \pa S/\pa Q^{\sfA}, t) = 0 \mbox{ } .
\label{HJE}
\ee

\ni This is to be solved as a p.d.e for the as-yet undetermined {\it Hamilton's principal function} $S(Q^{\sfA}, t)$.  
Then as a particular subcase, if the Hamiltonian is itself time-independent, one can use $S = \chi(Q^{\sfA}) - Et$ as a separation ansatz. 
This then obeys 
\beq
H(Q^{\sfA}, \pa \chi/\pa Q^{\sfA}) = E \mbox{ } .
\label{Stat-HJE}
\eeq 
This is to be solved for {\it Hamilton's characteristic function} $\chi$
As well as sometimes being computationally useful, 
the Hamilton--Jacobi formulation is close to the semiclassical approximation to QM and the Semiclassical Approach to the Problem of Time and Quantum Cosmology. 

\mbox{ } 

\ni [In the Routhian formulation, the constancy of $p^c_{\sfY}$ -- the sole momenta involved -- precludes an analogue of Hamilton--Jacobi theory prior to further Legendre transformations.
On the other hand, in the anti-Routhian formulation, 

\ni $\pa S/\pa t + A(Q^{\sfX}, \pa S/\pa Q^{\sfX}, \dot{c}^{\sfY}, t) = 0$ makes sense in the $t$-dependent case, 
and $A(Q^{\sfX}, \pa S/\pa Q^{\sfX}, \dot{c}^{\sfY}) = E$ in the $t$-independent case.
However the current Article in any case bypasses use of this last equation.]

\subsection{Hamiltonian formulation in the case with constraints}\label{Constraints}

Passage from Lagrangian to Hamiltonian formulation can be nontrivial. 
The {\it Legendre matrix}
\beq
\Lambda_{\sfA\sfB} := {\pa^2 L}/{\pa \dot{Q}^{\sfA}\pa\dot{Q}^{\sfB}} \mbox{ } \big( = {\pa P_{\sfB}}/{\pa \dot{Q}^{\sfA}} \big)  \mbox{ } 
\label{Leg-Matrix}
\eeq
-- named by its association with the Legendre transformation -- is in general non-invertible, so the momenta cannot be independent functions of the velocities.\footnote{In the case of 
the purely-quadratic kinetic term action, this is just $M_{\tfA\tfA^{\prime}}$.}
%
I.e. there are relations of the type ${\scC}_{\sfC}(Q^{\sfA}, P_{\sfB}) = 0$ between the momenta.  
This is quite a general type\footnote{I.e. these are scleronomous equality constraints,
as opposed to the thus-excluded {\it rheonomous} constraints.
Inequality constraints are also excluded.
Another term often used to describe constraints is {\it holonomic}, meaning that it is integrable into the form $f(Q^{\sfA}, t) = 0$.
Note that there are two ways this can fail: nonintegrability {\sl or} inequality constraints.
Now, whilst Dirac's treatment solely considers equality constraints, it says nothing about whether these are to be integrable or not, and hence are holonomic or the first sense of not so.}
type of constraint considered by Dirac.
Indeed, the Euler--Lagrange equations can be rearranged to reveal the explicit presence of the Legendre matrix,  
\beq
\ddot{Q}^{\sfA^{\prime}} \,  \pa^2 L/\pa \dot{Q}^{\sfA^{\prime}}  \pa\dot{Q}^{\sfA}  = \pa L/\pa Q^{\sfA} 
                                                                                     - \dot{Q}^{\sfA^{\prime}} \, \pa^2 L/\pa {Q}^{\sfA} \pa \dot{Q}^{\sfA^{\prime}} \mbox{ } .
\label{acc}
\eeq
By this, its noninvertibility has additional significance as accelerations not being uniquely determined by $Q^{\sfA}$, $\dot{Q}^{\sfA}$.\footnote{I remind the reader that this 
basic account of Physics restricts itself to no higher than second-order theories.} 

Constraints as arising from the above non-invertibility of the momentum--velocity relations are furthermore termed {\it primary}, 
as opposed to those requiring input from the variational equations of motion which are termed {\it secondary} \cite{Dirac, HTbook}.  
I index these by $\fP$ and $\fS$ respectively.
$\scH$ and $\scE$ illustrate that the primary-secondary distinction is artificial insofar as it is malleable by change of formalism. 

\mbox{ } 

\ni Since the Routhian and anti-Routhian also arise by Legendre transformations, one may wonder at this stage whether these involve their own notions of Legendre matrix. 
In the case of the Routhian, $\Lambda_{\sfY\sfY^{\prime}} := {\pa^2 L}/{\pa \dot{c}^{\sfY}\pa\dot{c}^{\sfY^{\prime}}} = 0$ by (\ref{cyclic-vel}), 
so it is an unintersting albeit entirely obstructive object.
The corresponding expressions for acceleration are similarly entirely free of reference to the cyclic variables.
On the other hand, the anti-Routhian's $\Lambda_{\sfX\sfX^{\prime}} := {\pa^2 L}/{\pa \dot{Q}^{\sfX}\pa\dot{Q}^{\sfX^{\prime}}}$ is in general nontrivial, 
and one can base a theory of primary constraints on it rather than on the usual larger (\ref{Leg-Matrix}).
The smaller anti-Routhian trick is then that the acceleration of $Q^{\sfX}$ is unaffected by the cyclic variables.
I.e. one can take (\ref{acc}) again with $\fX$ in place of $\fA$ since the further terms involving the cyclic variables that arise from the chain rule are annihilated by (\ref{cyclic-vel}).

\mbox{ }

\ni Dirac also introduced the concept of {\it weak equality} $\approx$, i.e. equality up to additive functionals of the constraints.
`Strong equality', on the other hand, means equality in the usual sense.  

\mbox{ }

\ni For $H$ one's incipient or `bare' Hamiltonian, one can additively append one's formalism's primary constraints using a priori any functions $f$ of the 
$\mbox{\boldmath $Q$}$ and $\mbox{\boldmath $P$}$ to form Dirac's further `starred' Hamiltonian $H^* := H + f^{\sfP}\scC_{\sfP}$.
Dirac's {\it total Hamiltonian} is $H_{\sT\so\st\sa\sll} := H + u^{\sfP}\scC_{\sfP}$, where the $u^{\sfP}$ are now regarded as unknowns.
One then begins to consider 
$\dot{\scC}_{\sfP} = \mbox{\bf \{} \scC_{\sfP} \mbox{\bf ,} \, H \mbox{\bf \}} + u^{\sfP^{\prime}} \mbox{\bf \{} \scC_{\sfP}  \mbox{\bf ,} \,  \scC_{\sfP^{\prime}} \mbox{\bf \}} \approx 0$
with the functions $u^{\sfP}$ now regarded as unknowns.

\mbox{ }

\ni The {\it Dirac algorithm} \cite{Dirac} then permits five kinds of outcomes.

\mbox{ } 

\ni Outcome 0) Inconsistencies.

\ni Outcome 1) Mere identities -- equations that reduce to $0 \approx 0$, i.e. $0 = 0$ modulo the $\scC_{\sfP}$.

\ni Outcome 2) Equations independent of the unknowns $u^{\sfP}$, which constitutes an extra constraint and is secondary.

\ni Outcome 3) Demonstration that existing constraints are in fact second-class (see below). 

\ni Outcome 4) Relations amongst the appending functions $u^{\sfP}$'s themselves, which are a further type of equation termed `{\it specifier equations}' 
(i.e. specifying restrictions on the $u^{\sfP}$).

\mbox{ }

\ni Lest 0) be unexpected, Dirac supplied a basic counterexample to PoD formulations entailing consistent theories.  
For $L = x$, the Euler--Lagrange equations read $0 = 1$. 

\mbox{ } 

\ni Suppose 2) occurs.  
Then defining `$\fQ = \fP + \fS$' indexing the constraints obtained so far, one can restart with a more general form form the problem,
\beq
\dot{\scC}_{\sfQ} = \mbox{\bf \{} \scC_{\sfQ} \mbox{\bf ,} \,  H \mbox{\bf \}} + u^{\sfP} \mbox{\bf \{} \scC_{\sfQ}  \mbox{\bf ,} \,  \scC_{\sfP} \mbox{\bf \}} \approx 0 \mbox{ } .
\label{de-for-u}
\eeq  
It may be necessary to do this multiple times [but clearly terminate if 0) occurs].
Suppose now that we have finished what solving there is to be made for the final problem's $u^{\sfP}$. 
These are of the form $u^{\sfP} = U^{\sfP} + V^{\sfP}$ by the split into i) $U^{\sfP}$ the particular solution.
ii) $V^{\sfP} = v^{\sfz}{V^{\sfP}}_{\sfZ}$ for $\fZ$ indexing the number of independent solutions in the `complementary function': 
the general solution of the corresponding homogeneous system
\beq
V^{\sfP} \mbox{\bf \{} \scC_{\sfC}  \mbox{\bf ,} \,  \scC_{\sfP} \mbox{\bf \}} \approx 0 
\eeq
with $v^{\sfA}$ the totally arbitrary coefficients of the independent solutions indexed by $\fA$.
This done, Dirac also defined the `primed Hamiltonian' $H^{\prime} := H +   U^{\sfP}\scC_{\sfP}$ 
These can be viewed as appendings by a determined mixture of free and fixed Lagrange multipliers.

A further classification of constraints is into {\it first-class} constraints (indexed by $\fF$ which weakly close under the classical bracket 
(ab initio the Poisson bracket, but this can change during the procedure).
{\it Second-class} constraints are then simply defined by exclusion as those constraints that fail to be first-class.  
First-class constraints use up 2 degrees of freedom each, and second-class constraints 1.

At least in the more standard theories of Physics, first-class secondary constraints arise from variation with respect to mathematically disjoint auxiliary variables.
Furthermore, the effect of this variation is to additionally use up part of an accompanying mathematically coherent block that however only contains partially physical information.

Some constraints are regarded as gauge constraints; however in general exactly which constraints these comprise remains disputed.
It is however agreed upon that second-class constraints are not gauge constraints; all gauge constraints use up two degrees of freedom.  
Dirac \cite{Dirac} conjectured a fortiori that all first-class constraints are gauge constraints,\footnote{This is in 
Dirac's sense of Gauge Theory \cite{DiracObs, Dirac}: concerning {\sl data at a given time}, so `gauge' here means {\it data-gauge}.
Contrast this with Bergmann's perspective \cite{Bergmann61} that that Gauge Theory concerns {\sl whole paths} (dynamical trajectories), so `gauge' there means {\it path-gauge}.}
so that using up two degrees of freedom would then conversely imply being a gauge constraint.
However, e.g. \cite{HTbook} contains a counterexample refuting this conjecture.  
One feature of Gauge Theory is an associated group of transformations that are held to be unphysical.  
The above-mentioned disjoint auxiliary variables are often in correspondence with such a group.      
{\it Gauge-fixing conditions} $\mF_{\sfX}$ may be applied to whatever Gauge Theory (though one requires the final answers to physical questions to be gauge-invariant).

Dirac finally defined $H_{\sE\sx\st\se\sn\sd\se\sd} := H + u^{\sfP}\scC_{\sfP} + u^{\sfS}\scC_{\sfS}$.  

\mbox{ } 

\ni Particularly with quantization in mind, first-class constraints tend to be relatively unproblematic, but second-class ones cause difficulties. 
Thus it is fortunate that there exist procedures for freeing one's theory of second-class constraints. 
There are two directions one can take: the Dirac bracket method excises them, whereas the effective method extends the phase space with further auxiliary variables 
so as to `gauge-unfix' second-class constraints into first-class ones.

\mbox{ } 

\ni Strategy A) Passage to the {\it Dirac brackets}.  
This replaces the incipient Poisson brackets with
\beq
\mbox{\bf \{}    F    \mbox{\bf ,} \, G    \mbox{\bf \}}\mbox{}^{\mbox{\bf *}} := \mbox{\bf \{}    F    \mbox{\bf ,} \, G    \mbox{\bf \}} - 
\mbox{\bf \{}    F    \mbox{\bf ,} \, \scC_{\sfI}    \mbox{\bf \}}  \mbox{\bf \{}    \scC_{\sfI}    \mbox{\bf ,} \, \scC_{\sfI^{\prime}}    \mbox{\bf \}}^{-1}   
\mbox{\bf \{}    \scC_{\sfI^{\prime}}    \mbox{\bf ,} \, G    \mbox{\bf \}} \mbox{ } . 
\eeq
Here the --1 denotes the inverse of the given matrix whose $\fI$ indices index irreducibly \cite{Dirac, HTbook} second-class constraints.
The classical brackets role played ab initio by the Poisson brackets is then taken over by the Dirac brackets.
The version of Dirac brackets formed once no second-class constraints remain illustrates the concept of `final classical brackets' forming a `final classical brackets algebra' 
of constraints.  

\ni Strategy B) Second-class constraints can always in principle\footnote{To this incipient statement of \cite{HTbook}, I add `locally', 
because gauge-fixing conditions themselves in general are not global entities.} 
be handled by alternatively thinking of them as `already-applied' gauge fixing conditions that can be recast as first-class constraints by adding suitable auxiliary variables.   
By doing this, a system with first- and second-class constraints {\it extends} to a more redundant description of a system with just first-class constraints. 

\mbox{ } 

\ni Moreover, Strategies A) and B) each make clear that the first class to second class distinction is also formalism-dependent.

\mbox{ } 

\ni So, all in all, one then enters the set of constraints in one's possession into the brackets in use to form a constraint algebraic structure.
This may enlarge one's set of constraints, or cause of one to adopt a distinct bracket. 
If inconsistency is evaded, the eventual output is an algebraic structure for all of a theory's constraints.

Symbolically, the algebraic structure formed by the constraints is 
\beq
\mbox{\bf \{} \scC_{\sfF} \mbox{\bf ,} \, \scC_{\sfF^{\prime}}\mbox{\bf \}}_{\sf\si\sn\sa\sll} = {C^{\sfF^{\prime\prime}}}_{\sfF\sfF^{\prime}}\scC_{\sfF^{\prime\prime}} \mbox{ } .  
\eeq
In some cases of relevance -- especially GR -- the ${C^{\sfF^{\prime\prime}}}_{\sfF\sfF^{\prime}}$ are structure functions rather than a Lie algebra's constants.

\subsection{Observables or beables}

It is then also natural to ask which quantities form zero brackets with a given closed algebraic structure of constraints, 
\beq
\mbox{\bf \{} \scC_{\sfC} \mbox{\bf ,} \, \iB_{\sfB}\mbox{\bf \}} \mbox{ } `=' 0 \mbox{ } 
\label{Beables}
\eeq
These entities are observables or beables\footnote{See \cite{ABeables} for discussion of the distinction between these two concepts.}, 
In the case of Dirac observables \cite{DiracObs} -- involving all the first-class constraints -- 
these entities are more useful than just any other function(al)s of the $Q^{\sfA}$ and $P_{\sfA}$ through containing solely physical information.
On the other hand, in the case of \K observables \cite{Kuchar93} involving all the first-class linear constraints,\footnote{These 
are often but not always the same \cite{ABeables} as the shuffle constraints of Sec \ref{CR-TRiPoD}.} 
these entities are of middling usefulness.  
This is due to being Dirac data gauge invariant quantities, albeit not ascertained to commute with the non-linear $\scC\scH\scR\scO\scN\scO\scS$ constraint as well.
Each such case itself closes as an algebraic structure associated with the corresponding constraint algebraic structure.
N.B. that for investigation of these notions of observables or beables, whichever of Dirac's Hamiltonians are equivalent. 
See e.g. \cite{Bergmann61, RovelliBook, ABeables} for yet further notions of observables or beables.

\subsection{Hamilton--Jacobi theory in the presence of constraints}\label{HJT-Constraints}

If the system has Dirac-type constraints, ${\scC}_{\sfC}(Q^{\sfA}; P_{\sfA} )$, the corresponding Hamilton--Jacobi equation 
-- whether (\ref{HJE}) or an {\it incipiently} timeless (\ref{Stat-HJE}) --- would be supplemented by 
\be
\scC_{\sfC}(Q^{\sfA}, \pa \chi/ \pa Q^{\sfA}) = 0 \mbox{ } .  \mbox{ }
\label{HJSUPPL} 
\ee

\section{Temporal Relationalism incorporating Principles of Dynamics (TriPoD)}\label{TRiPoD}

\subsection{Jacobi arc elements and Jacobi--Mach equations}

Consider now working in the absense of time at the primary level, as per Leibniz's Time Principle. 
[This is taken to include an absense of label time, as per Temporal Relationalism ii).]    
Restrict also for now to the finite second-order classical physical system.  
Now without time, there is no derivative with respect to time and thus no notion of velocity at the primary level.  
Thus one cannot use Lagrangian variables. 
However, Mach's Time Principle points to the availability of {\it change} 
at the primary level, so it is {\it Machian variables} $Q^{\sfA}, \d Q^{\sfA}$ that supplant the Lagrangian ones $Q^{\sfA}, \dot{Q}^{\sfA}$.
Now all information is contained within the {\it Jacobi arc element} $\d J(Q^{\sfA}, \d {Q}^{\sfA})$; 
this has supplanted the TRi time-independent Lagrangian $L(Q^{\sfA}, \dot{Q}^{\sfA})$ itself.
The action  $S$ is itself an {\sl unmodified} concept: it is already in TRiPoD form, 
albeit now additionally bearing the relation $S = \int \d J$ to the TRiPoD formulation specific Jacobi arc element $\d J$.
Actions of this form are manifestly parametrization irrelevant, thus indeed implementing Leibniz's Time Principle as per Sec 1.  
There is clearly also no primary notion of kinetic energy; this has been supplanted by {\it kinetic arc element} $\d s$ which for purely second-order systems is given by (\ref{S-Rel-2}).

Then $\d J = \sqrt{2W} \, \d s$ for $W(\mbox{\boldmath$Q$})$ the usual potential factor.  
In this manner, the kinetic and Jacobi arc elements are simply related by a conformal transformation;
In terms of $\d J$, dynamics has been cast in the form of a {\it geodesic principle} \cite{B94I}, 
whereas in terms of $\d s$ it has been cast in the form of a {\it parageodesic principle} \cite{Magic}.

Then apply the Calculus of Variations to obtain the equations of motion such that $S$ is stationary with respect to $Q^{\sfA}$.
The particular form that this variation takes is further commented upon in Sec \ref{FENoS}.  
Given the above set-up, I term the resulting equations of motion the {\it Jacobi--Mach equations}, 
\be
\d \left\{ \frac{\pa \, \d J}{\pa \, \d Q^{\sfA}} \right\} - \frac{\pa \, \d J}{\pa Q^{\sfA}} = 0   \mbox{ }. 
\label{JME}
\ee
These supplant the Euler--Lagrange equations.

\mbox{ }

\ni The Jacobi--Mach equations admit three corresponding simplified cases, as follows.

\mbox{ }

\ni 1) {\it Lagrange multiplier coordinates} $m^{\sfM} \, \underline{\subset} \, Q^{\sfA}$ are such that $\d J$ is independent of $\d m^{\sfM}$:
\beq
\pa \, \d J /\pa \, \d m^{\sfM} = 0 \mbox{ } .
\eeq
Then the corresponding Jacobi--Mach equation simplifies to 
\be
\pa \, \d J/\pa m^{\sfM} = 0 \mbox{ } .
\label{lmel-2}
\ee
2){\it Cyclic coordinates} $c^{\sfY} \, \underline{\subset} \, Q^{\sfA}$  are such that $\d J$ is independent of $c^{\sfM}$:
\beq
\pa \d J/\pa c^{\sfY} = 0 \mbox{ } ,
\eeq 
but features $\d c^{\sfY}$: the corresponding {\it cyclic differentials}.
Then the corresponding Jacobi--Mach equation simplifies to
\be
\pa \, \d J/\pa \, \d c^{\sfY} = \mbox{ const}_{\sfY} \mbox{ }.
\label{cyclic-vel-2}
\ee
3){\it The energy integral type simplification}.  
$\d J$ is independent of what was previously regarded as `the independent variable $t$', by which one Jacobi--Mach equation may be supplanted by the first integral
\be
\d J -  \frac{\pa \, \d J}{\pa \,\d Q^{\sfA}}\d Q^{\sfA} = \mbox{ constant } \mbox{ } .
\label{en-int-2}
\ee
Further suppose that 1)'s equations 
\beq
0 = \frac{\pa \, \d J}{\pa m^{\sfM}}(Q^{\sfO}, \d Q^{\sfO}, m^{\sfM}) \mbox{ happen to be solvable for } m^{\sfM} \mbox{ } .
\eeq
Then one can pass from $\d J(Q^{\sfO}, \d Q^{\sfO}, m^{\sfM})$ to a reduced $\d J_{\sr\se\sd}(Q^{\sfO}, \d {Q}^{\sfO})$: {\it Jacobi--Mach multiplier elimination}.

\mbox{ } 

\ni Note 1) More generally than for purely quadratic systems, the above formulation's presentation as a geodesic principle corresponding to some notion of geometry continues to hold 
-- for some more general notion of geometry (e.g. Finsler geometry). 
This more general case was pioneered by Synge (see \cite{Lanczos} for a discussion).  

\ni Note 2) Configuration--change space and configuration--velocity space are different presentations of the same tangent bundle $T($\FrQ$)$.

\ni Note 3) Formulation in terms of change $\d Q^{\sfA}$ can be viewed as introducing a {\it change covector}.
This is in the sense of inducing `{\it change weights}' to principles of dynamics entities, 
analogously to how introducing a conformal factor attaches conformal weights to tensorial entities.
For instance, $\d s$ and $\d J$ are change covectors too, whereas $S$ is a change scalar.  
Change scalars are entities which remain invariant under passing from PoD to TRiPoD, out of being already-TRi.

\ni Note 4) Finally, an explicit computation of (\ref{JME}) gives
\beq
\frac{\sqrt{2 W}\, \d}{||\d \mbox{\boldmath$Q$}||_{\mbox{\scriptsize \boldmath$M$}}}  
\left\{
\frac{\sqrt{2 W} \, \d Q^{\sfA}}{||\d \mbox{\boldmath$Q$}||_{\mbox{\scriptsize \boldmath$M$}}}  
\right\} 
+ \Gamma^{\sfA}\mbox{}_{\sB\sfC} \frac{\sqrt{2 W} \, \d Q^{\sfB}}{||\d  \mbox{\boldmath$Q$}||_{\mbox{\scriptsize \boldmath$M$}}}  
                                 \frac{\sqrt{2 W} \, \d Q^{\sfC}}{||\d  \mbox{\boldmath$Q$}||_{\mbox{\scriptsize \boldmath$M$}}}  = 
N^{\sfA\sfB}\frac{\pa W}{\pa Q^{\sfB}} \mbox{ } 
\label{New-Evol}
\eeq
for $\Gamma^{\sfA}\mbox{}_{\sB\sfC}$ the Christoffel symbols corresponding to $\mbox{\boldmath$M$}$.

\subsection{TRiPoD's formulation of momentum}

This is 
\be
P_{\sfA} := \pa \, \d J / \pa \,\d Q^{\sfA} \mbox{ } .  \label{TRiMom}
\ee
Furthermore, explicitly computing this for actions of type (2) gives the {\it momentum--change relation}
\be
P_{\sfA} = M_{\sfA\sfA^{\prime}} \frac{\sqrt{2W} \, \d Q^{\sfA^{\prime}} } { || \d {\mbox{\boldmath$Q$}}||_{\mbox{\scriptsize \boldmath$M$}}} \mbox{ } .
\label{New-P-Compute}
\ee
With (\ref{TRiMom}) depending on ratios of changes alone, momentum is a change scalar.

\subsection{TRiPoD's Legendre transformation}

One can now apply Legendre transformations that inter-convert changes $\d Q^{\sfA}$ and momenta $P_{\sfA}$.

\mbox{ } 

\ni Example 1) {\it Passage to the d-Routhian} 
\beq
\d R(Q^{\sfX}, \d Q^{\sfX}, P_{\scc}^{\sfY}) := \d J(Q^{\sfX}, \d Q^{\sfX}, \d c^{\sfY}) - P_{\sfY}^{\scc}\d c^{\sfY} \mbox{ } .
\eeq
{\it d-Routhian reduction} then requires being able to solve 

\ni\beq
\mbox{const}_{\sfY} = \frac{\pa \, \d J}{\pa \, \d c^{\sfY}}(Q^{\sfX}, \d Q^{\sfX}, \d c^{\sfY}) 
\eeq
as equations for the $\d c^{\sfY}$, followed by substitution into the above.
One application of this is the passage from Euler--Lagrange type actions to the geometrical form of the Jacobi actions without ever introducing a parameter.
See Sec \ref{CR-TRiPoD} for another application

\ni Example 2) {\it Passage to the d-anti-Routhian}.
\beq
\d A(Q^{\sfX}, P^{\sfX}, \d c^{\sfY}) = \d J(Q^{\sfX}, \d Q^{\sfX}, \d c^{\sfY}) - P_{\sfX}^{\scc}\d Q^{\sfX} \mbox{ } .
\eeq
A subcase of this plays a significant role in the next Sec.

\ni Example 3) {\it Passage to the d-Hamiltonian}.  
\beq
\d H(Q^{\sfA}, P_{\sfA}) =  P_{\sfA} \d Q^{\sfA} - \d J(Q^{\sfA}, \d Q^{\sfA}) \mbox{ } .  
\eeq
The corresponding equations of motion are in this case {\it d-Hamilton's equations} 
\beq
{\pa \, \d H}/{\pa P_{\sfA}} =  \d Q^{\sfA} \mbox{ } , \mbox{ } 
{\pa \, \d H}/{\pa Q^{\sfA}} = -\d P_{\sfA} \mbox{ } . 
\eeq

\subsection{The simpler case}

As per the Introduction, firstly $\scC\scH\scR\scO\scN\scO\scS$ arises as a primary constraint (this term is redefined in TRiPoD terms in Sec \ref{dA-Dir}, 
derivation cast in TRiPoD terms there too).
Secondly, $\scC\scH\scR\scO\scN\scO\scS$ can be rearranged to obtain expression (\ref{t-em-J}) for emergent time.
Also note that (\ref{en-int}) is the Jacobi--Mach formulation's manifestation of $\scC\scH\scR\scO\scN\scO\scS$: an equation of time rather than a statement of energy conservation.

Furthermore, adopting emergent Jacobi time simplifies the expression (\ref{New-P-Compute}) for the momenta to 
\beq
P_{\sfA} = M_{\sfA\sfB}\Last Q^{\sfB} \mbox{ } , 
\eeq
and the form of the Jacobi--Mach equations of motion (\ref{New-Evol}) to 
\beq
\Last\Last  Q^{\sfA} + \Gamma^{\sfA}\mbox{}_{\sfB\sfC}\Last Q^{\sfB}\Last Q^{\sfC} = N^{\sfA\sfB}\pa W/\pa Q^{\sfA} 
\label{parag-2} 
\mbox{ } .
\eeq 
The emergent Jacobi time thus has the desirable property of being a time with respect to which the equations of motion take a particularly simple form.
In this manner, in the case of Mechanics the emergent Jacobi time is a recovery of the same entity that is usually modelled by Newtonian time but now on relational premises.  
In the GR case, it amounts to a recovery of GR proper time (including of cosmic time in the cosmological case) \cite{ABook}.

\subsection{Configurational Relationalism and its TRiPoD form}\label{CR-TRiPoD}

\ni The corresponding harder case involves {\it Configurational Relationalism} as well.
This second aspect of Background Independence covers both of the following.  

\ni a) {\it Spatial Relationalism} \cite{BB82}: no absolute space properties.

\ni b) {\it Internal Relationalism}: the post-Machian addition of not ascribing any absolute properties to any additional internal space that is associated with the matter fields.

\ni These are notably different cases because b) holds at a fixed spatial point whereas a) move spatial points around.  

\ni Configurational Relationalism is then axiomatized as follows.

\mbox{ } 

\ni i) One is to include no appended configurational structures either (spatial or internal-spatial metric geometry variables that are fixed-background rather than dynamical).

\ni ii) Physics in general involves not only a $\FrQ$ but also a $\FrG$ of transformations acting upon $\FrQ$ that are taken to be physically redundant.

\mbox{ } 

\ni The usual Jacobi action principle itself is temporally-relational but spatially-absolute mechanics.     
On the other hand, Barbour and Bertotti \cite{BB82} found a means of freeing Mechanics actions from absolute space.
This involved using not $\dot{q}^{iI}$ but $\dot{q}^{iI} - A^i - \{\underline{B} \cr \underline{q}^I\}^i$ 
which allows for a frame that is translating and rotating in an arbitrary time-dependent manner.  
Unfortunately, these corrections ruin Manifest Reparametrization Invariance of the Jacobi action.
However \cite{ABFO, FEPI} the two implementations can be stacked together by use instead of $\dot{q}^{iI} - \dot{a}^i - \{\underline{\dot{b}} \cr \underline{q}^I\}^i$.
One can then pass to a Jacobi--Mach version of this that stacks Configurational Relationalism with 
Temporal Relationalism's geometric implementation that is dual to Manifest Parametrization Irrelevance \cite{FileR, ARel}. 
Here one uses $\d q^{iI} - \d a^i - \{\d \underline{b} \cr \underline{q}^I\}^i$.
These stackings are moreover nontrivial due to requiring careful considerations of how to vary auxiliary variables that are not just Lagrange multipliers, as per Sec \ref{FENoS}.  
The above suite of arbitrary frame corrections bear relation to gauge theory in Dirac's data-gauge sense.  
Let us furthermore pass to the more general case of $\d_gQ^{\sfA} := \d Q^{\sfA} - \stackrel{\rightarrow}{\FrG}_{\d g}Q^{\sfA}$, 
where $\stackrel{\rightarrow}{\FrG}_{\d g}$ denotes an infinitesimal group action of a group $\FrG$ acting on $\FrQ$, which action is held to be physically irrelevant.
In particular, the GR case of this then involves the spatial diffeomorphisms (and so is trivial for minisuperspace).  
Electromagnetism, Yang--Mills theory and corresponding gauge theories gauge symmetries can be considered along the lines of the internal part of Configurational Relationalism.

\mbox{ }

\ni The finite cases' relational action is then
\beq
S = \sqrt{2}\int \d s_g \sqrt{W} \mbox{ } , \mbox{ }  \mbox{ }  \d s_g := ||\d_g\mbox{\boldmath$Q$}||_{\mbox{\scriptsize \boldmath$M$}} \mbox{ } .
\label{S-Rel-G}
\eeq
\mbox{ } \mbox{ } The conjugate momenta are now
\be
 P_{\sfA} := \frac{\pa  \, \d J }{ \pa  \, \d{Q}^{\sfA}}  =  M_{\sfA\sfB} \, \frac{\sqrt{2W}}{||\d_g\mbox{\boldmath$Q$}||_{\mbox{\scriptsize \boldmath$M$}}} \, \d_{g}  Q^{\sfA} \mbox{ } .  
\ee
These obey one primary constraint per relevant notion of space point, interpreted as an equation of time (\ref{CHRONOS}), so it is purely quadratic in the momenta.
They also obey $\fG$ secondary constraints per relevant notion of space point from variation with respect to $g^{\sfG}$  
\beq
0 = \pa \, \d J/\pa  \, \d c^{\sfG} = P_{\sfA} \, \mbox{$\delta$} \, \d g^{\sfG} / \mbox{$\delta$} \{\stackrel{\rightarrow}{\FrG}_{\d g} Q^{\sfA}\} 
                                              := \scS\scH\scU\scF\scF\scL\scE_{\sfG} \mbox{ } ;     
\label{LinZ}
\eeq
also note that these are linear in the momenta.
Here also $\delta$ denotes variational derivative.

Next, denote the joint set of these constraints by $\scC_{\sfF}$, under the presumption that they are confirmed as first-class as per Sec \ref{dA-Dir}.
The indexing set designation assumes there is only one quadratic constraint, so all our examples' $\fF$ ranges over $\fG$ plus one value indexing $\scC\scH\scR\scO\scN\scO\scS$.

The arbitrary $\FrG$, arbitrary $\FrQ$ generalization of Barbour and Bertotti's {\it Best Matching} method -- also reformulated in Jacobi--Mach terms -- 
then consists of the following steps.

\mbox{ }

\ni Best Matching 1) Start with the `arbitrary $\FrG$ frame corrected' action (\ref{S-Rel-G}).  

\ni Best Matching 2) Next, extremizes over $\FrG$ in accord with the variational principle including variation with respect to $\d  g^{\sfG}$.
This extremization produces the above linear constraint equation $\scL\scI\scN_{\sfG} = 0$.  

\ni Best Matching 3) In Machian variables $ Q^{\sfA}, \d { Q}^{\sfA}$, this is to be solved for the $\d g^{\sfG}$ themselves.

\ni Best Matching 4) This solution is then to be substituted back into the action, in a further example of d-Routhian reduction. 
This produces a final $\FrG$-independent expression. 

\ni Best Matching 5) Elevate this new action to be one's primary starting point.  

\mbox{ } 

\ni A further step, added later with gradually increasing detail \cite{B94I, FileR, ARel, AM13} is that the classical Machian emergent time now takes the form
\beq
t^{\se\sm(\sJ\sB\sB)} -  t^{\se\sm(\sJ\sB\sB)}(0)                          = \mbox{\large E}_{\d g \mbox{ } \in \mbox{ } \sFG}
\left(                                                              
\int  {||\d_g\mbox{\boldmath$Q$}||_{\mbox{\scriptsize \boldmath$M$}}}/\sqrt{2 W(\mbox{\boldmath$Q$})} 
\right)  \mbox{ }  .  
\label{Kronos2}
\eeq
[JBB here stands for `Jacobi--Barbour--Bertotti' and E denotes extremum.] 
This illustrates that, in the absense of being able to explicitly solve the Best Matching, one does not have an explicit solution for the classical Machian emergent time either
On the other hand, if one succeeds in carrying out Best Matching.  
Then $\d  g$ is replaced by an extremal expression solely in terms of $\widetilde{Q}^{\sfR}, \d  \widetilde{Q}^{\sfR}$ with tildes usedto denote the reduced space quantities.  
By this, $t^{\se\sm(\sJ\sB\sB)}$ is expressed in terms of the {\sl reduced} configuration space's geometry: the tilded version of (\ref{t-em-J}). 
The latter occurs nontrivially {\sl and} solvably for all scaled 2-$d$ RPM's \cite{FileR} (simplest of which is the scaled relational triangle: Arena 5), 
as well as for pure-shape (i.e. scale free) RPM's in both 1- and 2-$d$. 
On the other had, the former occurs for full GR: Wheeler's Thin Sandwich Problem \cite{BSW}, which is listed as a second PoT facet \cite{Kuchar92I93}.
Finally note the title of \cite{BSW}: ``{\it three-dimensional geometry as carrier of information about time}. 
This clearly illustrates that Wheeler was already aware then of the time connotations of the thin sandwich subcase of the above `Best Matching' working. 
This title is furthermore a clear illustration of `duality between geometry and Manifest Parametrization Irrelevance' as an implementation of timelessness at 
the primary level leading to time being abstracted from change at the secondary level.

\subsection{Free endpoint variation}\label{FENoS}

Suppose a formulation's auxiliary multiplier coordinate $m$ is replaced by a cyclic velocity $c$ \cite{ABFO, FEPI} or cyclic differential $\d c$ \cite{FileR, ARel}. 
Then right-hand side zero of the multiplier equation is replaced by a constant in the corresponding cyclic equation.  
However, if the quantity being replaced is an entirely physically meaningless auxiliary, the meaninglessness of its values at the endpoint becomes nontrivial in the cyclic case. 
I.e. the appropriate type of variation is {\it free end point (FEP)} variation alias {\it variation with natural boundary conditions} \cite{FoxCHBrMa, Lanczos}.\footnote{To be clear,
this here refers to free value {\sl at} the end-NoS rather than the also quite commonly encountered freedom {\sl of} the end-NoS itself.} 
By this means it has enough freedom to impose more conditions than the more usual fixed-end variation does.
In particular, it imposes 3 conditions per variation, 
\beq
{\pa \, \d J}/{\pa g^{\sfG}} = \d{ p}_{\sfG} \mbox{ } , \mbox{ } \mbox{ alongside } \mbox{ }  
\left.
 p_{\sfG}
\right|_{\mbox{\scriptsize endpoint $a$}} = 0 \mbox{ } , \mbox{ } \mbox{ } a = 1, 2 \mbox{ } . 
\label{correct}
\eeq
Case 1) If the auxiliaries $g^{\sfG}$ are multipliers $m^{\sfG}$, (\ref{correct}) reduces to
$
p_{\sfG} = 0 \mbox{ } , \mbox{ } {\pa L}/{\pa  m^{\sfG}} = 0
$
and redundant equations.
Thus the endpoint value terms automatically vanish in this case by applying the multiplier equation to the first factor of each.  
This is the case regardless of whether the multiplier is not auxiliary and thus standardly varied, or auxiliary and thus FEP varied. 
This is because this difference in status merely translates to whether or not the cofactors of the above zero factors are themselves zero or not.  
Thus the FEP subtlety in no way affects the outcome in the multiplier coordinate case.  
Thus the FeP subtlety remained unnoticed whilst following Dirac and ADM in encoding fundamental physics auxiliaries as multiplier coordinates.

\ni Case 2) If the auxiliaries $ g^{\sfG}$ are considered to be cyclic coordinates $c^{\sfG}$, (\ref{correct}) reduces to 
\beq 
\left.
p_{\sfG}
\right|_{\mbox{\scriptsize endpoint a}} = 0 \mbox{ } 
\label{ckill}
\eeq
alongside 
\beq 
\d{ p}_{\sfG} = 0 \mbox{ } \Rightarrow \mbox{ } \mbox{ } p_{\sfG} = C \mbox{ } . 
\label{hex}
\eeq 
Then $C$ is identified as 0 at either of the two endpoints (\ref{ckill}).
Being invariant along the curve of notions of space, it is therefore zero everywhere.  
Thus (\ref{hex}) and the definition of momentum give 
\beq
\pa \, \d \, J/\pa \d{ c}^{\sfG} = 0 \mbox{ } .  
\eeq
In conclusion, the above FEP working ensures that the cyclic and multiplier formulations of auxiliaries in fact give {\sl the same} variational equation. 
Thus complying with Temporal Relationalism by passing from encoding one's {\sl auxiliaries} as multipliers to encoding them as cyclic velocities or differentials 
is valid without spoiling the familiar and valid physical equations.

Note that a similar working \cite{FEPI} establishes that for an auxiliary formulated in the cyclic manner, 
passage to the (d-)Routhian reproduces the outcome of multiplier elimination in the case of that same auxiliary entity being formulated as a multiplier.

\subsection{TRi-morphisms and brackets. i.}

Suppose there are no cyclic differentials to be kept.
Then $\FrQ$ morphisms carry down, except that specifically Point rather than Point$_t$ is involved.
The Liouville 1-form (\ref{Liouville}) is already TRi and thus a change 1-form, whereas the symplectic 2-form (\ref{Symplectic}) is already TRi and thus a change 2-form.
The use of Can rather than Can$_t$ is also now imperative in the d-Hamiltonian formulation.  
Also Phase space carries down since all of $\Q^{\sfA}$, $\P_{\sfA}$ and the Poisson bracket carry over.

\subsection{dA-Hamiltonians, -phase space and -Dirac algorithms}\label{dA-Dir}

The Legendre matrix encoding the non-invertibility of the momentum-velocity relations is now supplanted by the 

\ni{\it $\d^{-1}$-Legendre matrix}
\beq
\d^{-1} \Lambda_{\sfA\sfB} := {\pa^2 \d s}/{\pa \, \d Q^{\sfA} \pa \, \d Q^{\sfB}} \mbox{ } \left( = {\pa P_{\sfB}}/{\pa \, \d Q^{\sfA}} \right)  \mbox{ } 
\eeq
encoding the non-invertibility of the momentum--change relation.
N.B. here $\d^{-1}$ denotes that this is a change {\sl vector}.
Then the TRiPoD definition of primary constraint follows from this in parallel to how the usual definition of primary constraint follows from the Legendre matrix, 
with secondary constraint remaining defined by exclusion of primary constraints.

\mbox{ } 

\ni Example 1) Dirac's argument for Reparametrization Invariance implying at least one primary constraint then takes the following TRiPoD form.
A geometrical action is dually parametrization-irrelevant. 
Thus it is homogeneous of degree 1 in its changes. 
So each of its total of $k$ momenta are homogeneous of degree 0 in the changes.
Thus these are functions of $k$ -- 1 independent ratios of changes. 
So there must be at least one relation between the momenta themselves without any use made of the equations of motion.  
But by definition, this is a primary constraint.

Then the specific form of the primary constraint          that follows from actions of the form (\ref{S-Rel-2}) 
is of course the same $\scC\scH\scR\scO\scN\scO\scS$ that follows from actions of the form (\ref{S-Rel}).

\mbox{ } 

\ni The next idea in building a TRiPoD version of Dirac's general treatment of constraints is to append constraints to one's incipient d-Hamiltonian not with Lagrange multipliers 
(which would break TRi) but rather with cyclic differentials.
Thus a {\it dA-Hamiltonian} is formed; `A' here stands for `almost'. 
The dA-Hamiltonian is a particular case of d-anti-Routhian. 
Note moreover that the dA-Hamiltonian $\d A$ symbol has an extra minus sign relative to the d-anti-Routhian $\d A$ symbol.
This originates from the definition of Hamiltonian involving an overall minus sign where the definitions of Routhian and anti-Routhian have none.  
And which particular case?
The one in which all the cyclic coordinates involved have auxiliary status and occur in best-matched combinations.
[In the event of a system possessing physical as well as auxiliary cyclic velocities this would be a `partial' rather than `complete' Routhian, 
though this case does not further enter this Article.]

The equations of motion are then {\it dA-Hamilton's equations} are 
\beq
{\pa \, \d A}/{\pa \, P_{\sfA}} =  \dot{Q}^{\sfA} \mbox{ } , \mbox{ } 
{\pa \, \d A}/{\pa \, Q^{\sfA}} = -\dot{P}_{\sfA} \mbox{ } , 
\label{dA-Ham-eqs}
\eeq
augmented by $\pa \, \d A / \pa \, \d c^{\sfG} =  0$.
Also note that Sec 2.7's comment about using the anti-Routhian's own Legendre matrix carries over to the d-anti-Routhian 
and thus also to the reidentification of a subcase of this as the dA-Hamiltonian.

\mbox{ }

\ni Examples of the above appendings, paralleling Dirac's treatment, are then firstly the {\it starred dA-Hamiltonian} 
$\d A^* := \d A + \d f^{\sfP} \scC_{\sfP}$ for arbitrary functions of $Q^{\sfA}, P_{\sfA}$ now represented as cyclic differentials $\d f(Q^{\sfA}, P_{\sfA})$.
Secondly, the {\it total dA-Hamiltonian} $\d A_{\sT\so\st\sa\sll} := \d A + \d u^{\sfP} \scC_{\sfP}$, where $\d u^{\sfP}$ are now unknown cyclic differentials.

The next issue to arise is the counterpart of the bracket expression in (\ref{de-for-u}).
Now the best-matched form of the action ensures the constraints are of the form $\sfC(Q^{\sfA}, P_{\sfA} \mbox{ alone})$, because for these 
\beq
\mbox{passage from $\d Q^{\sfA}$ to $P_{\sfA}$ absorbs all the $\d g^{\sfG}$} \mbox{ } .
\label{key}
\eeq

In the case in hand, $\scC^{\scC} = \scC(Q^{\sfA}, P_{\sfA} \mbox{ alone})$ means that the chain-rule expansion 
$$
\d \scC_{\sfC} = \frac{\pa \, \d \scC_{\sfC}}{\pa \, Q^{\sfA}} \, \d Q^{\sfA} + \frac{\pa \, \d \scC_{\sfC}}{\pa P^{\sfA}} \, \d P^{\sfA}
$$
applies, whence by (\ref{dA-Ham-eqs})
\beq
\d\scC_{\sfC} = \mbox{\bf \{} \scC_{\sfC} \mbox{\bf ,} \, \d A_{\sT\so\st\sa\sll} \mbox{\bf \}} = 
                \mbox{\bf \{} \scC_{\sfC} \mbox{\bf ,}    \d A                    \mbox{\bf \}} + \d u^{\sfP} \mbox{\bf \{} \scC_{\sfC} \mbox{\bf ,} \, \scC_{\sfP} \mbox{\bf \}} \mbox{ } . 
\eeq
\mbox{ } \mbox{ }  Next let this be solved for unknown cyclic differentials under the split $\d u^{\sfP} = \d U^{\sfP} + \d V^{\sfP}$ in direct parallel with Dirac's: 
a cyclic differential particular solution $\d U^{\sfP}$ and a cyclic differential complementary function $\d V^{\sfP} = v^{\sfZ}\d{V^{\sfP}}_{\sfZ}$.
Then the {\it primed dA-Hamiltonian} $\d A^{\prime} := \d A + \d U^{\sfP} \scC_{\sfP}$ where $\d U^{\sfP}$ are now the cyclic differential particular solution part of $\d u^{\sfP}$
Finally, the {\it extended dA-Hamiltonian} $\d A_{\sE\sx\st\se\sn\sd\se\sd} := \d A + \d u^{\sfP} \scC_{\sfP} + \d u^{\sfS} \scC_{\sfS}$.

\mbox{ } 

\ni Fig \ref{TRiPoD-Squares}.b) exposits the significance via the usual cyclic velocity intermediate in Fig \ref{TRiPoD-Squares}.a).  

\ni Phase is then replaced by  A-Phase and $\d$A-Phase; these are all types of bundle twice over (cotangent bundles {\sl and} $\FrG$ bundles).
 
{            \begin{figure}[ht]
\centering
\includegraphics[width=1.0\textwidth]{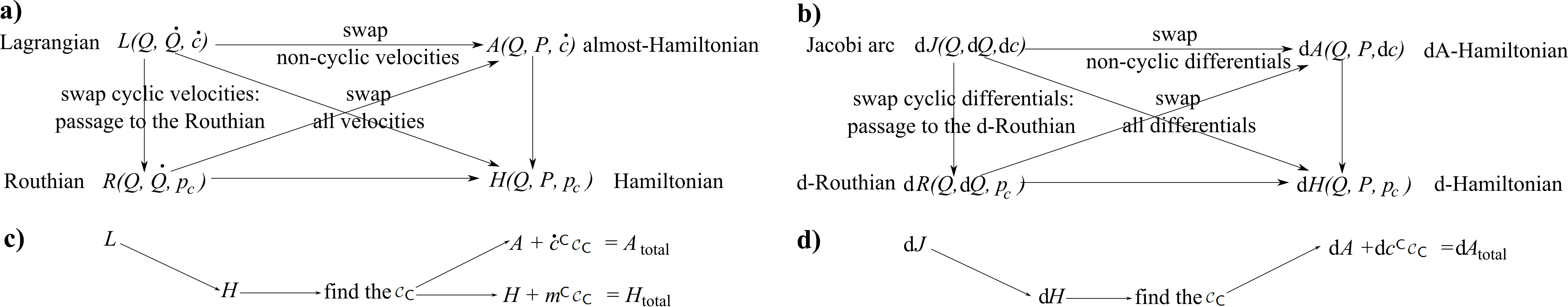}
\caption[Text der im Bilderverzeichnis auftaucht]{         \footnotesize{a) Almost-Hamiltonian subcase of Fig 1.a)'s Legendre square.    
b) then elevates this to fully TRiPoD form in terms of $\d$-Legendre transformations that dually switch momenta and changes.  
These are between change covectors: d$J$, d$R$, d$A$, d$H$, the information-preserving extra terms now being subsystem Liouville forms, which were always change covectors. 
Routhians go to $\d$-Routhians 
(there is no need for `almost' in this case since Routhians are already allowed to contain velocities, and so already include almost-Routhians as a subset). 
c) and d) exhibit the choices by which the total Hamiltonian, A-Hamiltonian and dA-Hamiltonian arise. 
The starred, primed and extended versions follow suit.} }
\label{TRiPoD-Squares}\end{figure}            }

\ni For example, in Arena 5: the scaled relational triangle, $\d A_{\sT\so\st\sa\sll} := \d I \, \scE   +   \d A^i \scP_i + \d B^i \scL_i $  (strictly $ + \d\mu^i P^A_i + \d\nu^i P^B_i$).
Here $P_i$ is the zero total momentum constraint and $L_i$ is the zero total angular momentum constraint.
$\d I$ is the differential of the instant, which is numerically equal to $\d t$ for Mechanics, albeit obtained from a relational perspective rather than assumed.  
On the other hand, there is no d-Hamiltonian to dA-Hamiltonian distinction for minisuperspace.  

\mbox{ }

\ni From only the Poisson bracket part acting on the constraints, it follows that the definition of first-class and second class remain unaffected. 
So are the Dirac bracket and the extension procedure.

\mbox{ }

\ni Also Dirac's algorithm is now supplanted by the $\d$A-Dirac algorithm.
The five cases this is capable of producing at each step, in any combination, are as follows.  

\mbox{ } 

\ni Outcome 0), 1) and 3) are as before.

\ni Outcome 2) Equations independent of the unknowns $\d u^{\sfP}$, which constitute an extra constraint and is secondary.
[This independence perpetuates the Poisson bracket sufficing for action upon constraints.]  

\ni Outcome 4) Relations amongst the appending functions $\d u^{\sfP}$ themselves: specifier equations now specifying restrictions on the formulation's auxiliary cyclic differentials.

\mbox{ }  

\ni The algebraic structure of the constraints is unaffected by passing to TRiPoD form: same constraints and same brackets for the purpose of acting on the constraints.  
This is other than a minor and physically inconsequential point of difference in formulation of the smearing in the field-theoretic case.

\subsection{TRi-morphisms and brackets. ii.}

Suppose there are now cyclic ordials to be kept, or arising from the dA-Dirac procedure.
Then the morphisms are a priori of the mixed type Can($T^*(\FrQ)$) $\times$ Point($\FrG$).
Also the brackets are a priori of the mixed Poisson--Peierls type: Poisson as regards $Q^{\sfA}, P_{\sfA}$ and Peierls as regards $d g^{\sfG}$. 

\mbox{ } 

\ni Moreover, (\ref{key}) has the following implications.

\ni i) Can($T^*(\FrQ)$) $\times$ Point($\FrG$) act as just Can($T^*(\FrQ)$) on the constraints.

\ni ii) As far as the constraints are concerned, the brackets act as just Poisson brackets on $Q^{\sfA}, P_{\sfA}$.   
{\sl The physical part of the dA-Hamiltonian's incipient bracket is just a familiar Poisson bracket}.  
This good fortune follows from the dA-Hamiltonian being a type of d-anti-Routhian, 
alongside its non-Hamiltonian variables absenting themselves from the constraints due to the best-matched form of the action.

\subsection{TRi observables or beables}

Much as how for investigation of these notions of observables or beables, whichever of Dirac's Hamiltonians are equivalent, the same applies to A- and dA-variants.
This includes definitions based on cases of (\ref{Beables}) as well as the particular PDEs to solve for specific examples, 
since those are based on the same constraints and the same brackets.
As per the end of the previous Subsec, there remains a minor point of difference in formalism of the smearing in the field-theoretic case.

\mbox{ }  

\ni \K observables are further motivated in the relational approach by $\scS\scH\scU\scF\scF\scL\scE$ and $\scC\scH\scR\scO\scN\scO\scS$ arising separately.  
See \cite{ABeables, FileR} for a number of examples of \K observables, including for arenas 2 and 5.

\subsection{TRiPoD's Hamilton--Jacobi Theory}

This is a $t$-independent version, for a totally constrained dA-Hamiltonian.
In such a case, only the constraints themselves feature. 
Then (\ref{key}) implies that the form of this is no different from the usual PoD's corresponding $t$-independent totally constrained Hamiltonian case. 
Thus the closeness to the Semiclassical Approach carries over from the standard PoD case to here.

The Hamilton--Jacobi formulation of relational theories thus consists of 
\beq
\scC\scH\scR\scO\scN\scO\scS(Q^{\sfA}, \pa \chi/\pa Q^{\sfA}) = 0 \mbox{ } ,
\eeq 
with allowed extra dependence on a meaningful constant such as $E$ for Mechanics or $\Lambda$ for (minisuperspace) GR.
N.B. Hamilton's characteristic function $\chi$ is a change scalar.  
In the case of nontrivial $\FrG$, this is supplemented by
\beq
\scS\scH\scU\scF\scF\scL\scE_g(Q^{\sfX}, \pa \chi/\pa Q^{\sfX}) = 0 \mbox{ } .
\eeq

\section{Conclusion}\label{Conclusion}

\subsection{Three-legged summary}

\ni 
See Fig \ref{TRiPoD6} for a summary of which Principles of Dynamics (PoD) structures and formulations require supplanting in order to be 
Temporal Relationalism implementing (TRi) and which are already satisfactory.
The red leg is supplanted by the green leg whilst the blue leg is unaffected: the ``already temporally relational" parts of the standard PoD.
This mostly consists of change scalars.
The new green leg is powered by the FENoS (free end notion of space) variation of Sec \ref{FENoS}). 
The differential almost Hamiltonian formulation is as good at handling classical constraints as the Hamiltonian formulation itself, and bridges just as well to quantum theory. 
Its advantage is that it is TRi as well, so it has extra applicability to whole universes and other closed models.  
If the d$g^{\sfG}$ can be reduced out, many of the differences in the above figure collapse down to their forms: 
all A's are lost, leaving a lesser number of entities still possessing a distinct d form.  
Finally, explicitly time-dependent structures such as time-dependent Lagrangians, Hamiltonians, point transformations and canonical transformations have no TRiPoD counterparts.

{            \begin{figure}[ht]
\centering
\includegraphics[width=0.72\textwidth]{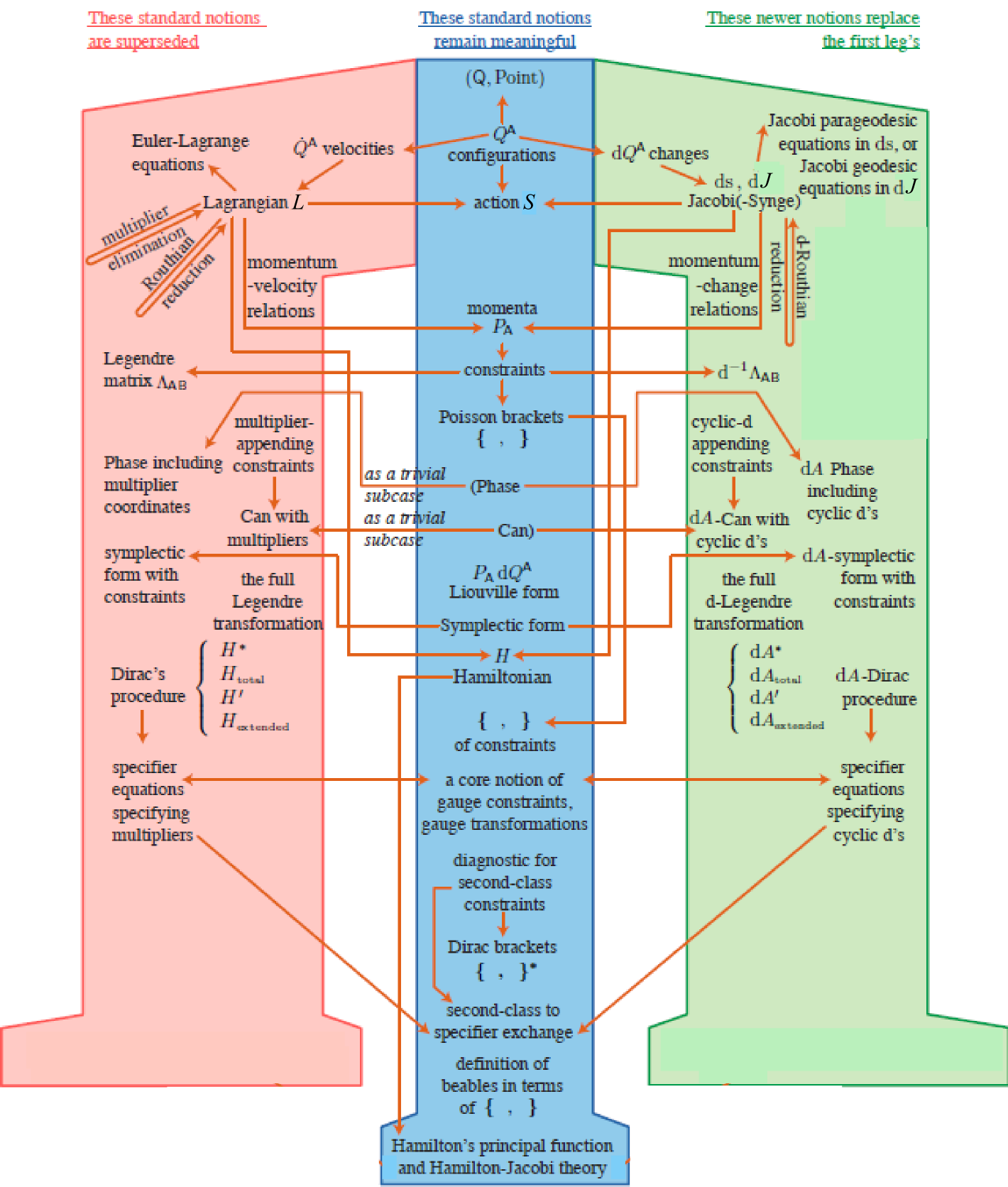}
\caption[Text der im Bilderverzeichnis auftaucht]{        \footnotesize{The Temporal Relationalism incorporating Principles of Dynamics (TRiPoD). 
The powers of d displayed indicate the change tensoriality of each entity. 
`A' stands for `almost', meaning that auxiliary entities are treated differently but all physical entities are still treated the same way.
Can denotes canonical transformations. } }
\label{TRiPoD6}\end{figure}            }

\subsection{Further extensions of TRi}

\ni 1) In the further Spacetime Construction working \cite{AM13} (based in part on the earlier \cite{RWR}), the Dirac Algorithm gives not only Constraint Closure but also the 
form of GR's Hamiltonian constraint $\scH$, construction of GR-type spacetime and local Special Relativity.

\mbox{ } 

\ni Then one has to consider the following two extra facets of the PoT.

\mbox{ } 

\ni i) Spacetime Relationalism (of the unsplit but emergent spacetime) as per usual. 
There are to be no appended spacetime structures, in particular no indefinite background spacetime metrics. 
Fixed background spacetime metrics are also more well-known than fixed background space metrics. 
As well as considering a spacetime manifold $\FrM$, consider also a $\FrG_{\sS}$ of transformations acting upon $\FrM$ that are taken to be physically redundant.
For GR,  $\FrG_{\sS}$ = Diff($\FrM$).

\mbox{ } 

\ni ii) Foliation Independence of the split spacetime, for which it is now natural to consider a TRi version \cite{TRiFol} using the TRi A-split rather than the ADM-split. 
This leads to a new Machian interpretation \cite{AM13, TRiFol} of GR's Thin Sandwich \cite{BSW} and of \K's universal kinematics \cite{Kuchar76I-II}.

\mbox{ }  

\ni 2) One can also then consider TRi canonical quantum mechanics (TRiCQM) \cite{ABook} e.g. along the lines of geometrical quantization \cite{I84}. 
One can then apply semiclassical approximations to pass to the intermediate case of TRi semiclassical quantum cosmology (TRiSQC).


\end{document}